\documentclass{emulateapj}
\usepackage{apjfonts}
\def\hi{H\,{\sc i}}

\def\planck {\emph{Planck}}
\def\spitzer {\emph{Spitzer}}
\slugcomment{Submitted to AJ}

\shorttitle{Searching for dust in compact HVCs}
\shortauthors{Williams et al.}

\begin{document}

\title{A Spitzer-MIPS search for dust in compact high-velocity \hi\ 
clouds}

\author{Rik J. Williams\altaffilmark{1}, 
	Smita Mathur\altaffilmark{2,3},
	Shawn Poindexter\altaffilmark{2},
       Martin Elvis\altaffilmark{4},
        Fabrizio Nicastro\altaffilmark{5,4}}
\altaffiltext{1}{Carnegie Observatories, 813 Santa Barbara St., Pasadena, 
CA 91101, USA}
\altaffiltext{2}{Department of Astronomy, The Ohio State University,
                 Columbus, OH 43210, USA}
\altaffiltext{3}{Center for Cosmology and AstroParticle Physics, The
		 Ohio State University, Columbus, OH 43210, USA}
\altaffiltext{4}{Harvard-Smithsonian Center for Astrophysics, 60 Garden
                 St., Cambridge, MA 02138, USA}
\altaffiltext{5}{Osservatorio Astronomico di Roma-INAF, Via di Frascati 33, 
                 00040 Monte Porzio Catone, RM, Italy}
\email{williams@obs.carnegiescience.edu}

\begin{abstract}
We employ three-band Spitzer-MIPS observations to search for cold dust 
emission in three neutral hydrogen compact high-velocity clouds (CHVCs) in 
the vicinity of the Milky Way.  Far-infrared emission correlated with
\hi\ column density was previously reported in HVC Complex C,
indicating that this object contains dust heated by the Galactic
radiation field at its distance of $\sim 10$\,kpc.  Assuming published
Spitzer, IRAS, and Planck, IR-\hi correlations for Complex C, our \emph{Spitzer} 
observations are of sufficient depth to directly detect $160\mu$m dust 
emission in the CHVCs if it is present
at the same level as in Complex C, but no emission is detected in any
of the targets.  For one of the targets (CHVC289) which has well-localized \hi\ 
clumps, we therefore conclude that it is fundamentally different from 
Complex C, with either a lower dust-to-gas ratio or a greater distance 
from the Galactic disk (and consequently cooler dust temperature).  Firm 
conclusions cannot be drawn for the other two
\emph{Spitzer}-observed CHVCs since their small-scale \hi\ structures
are not sufficiently well known; nonetheless, no extended dust emission
is apparent despite their relatively high \hi\ column densities.
The lack of dust emission in CHVC289 suggests that at least some compact
high-velocity clouds objects may exhibit very low dust-to-gas ratios
and/or greater Galactocentric distances than large HVC complexes.
\end{abstract}


% Vega-AB conversion: 1.900 (K), 0.938 (J)

\section{Introduction}
Neutral hydrogen high-velocity clouds (\hi\ HVCs) have been known to exist for
nearly half a century \citep{muller63}, but their cosmological significance and
role in the Milky Way and Local Group remain ambiguous.  HVCs span a wide range
of sizes on the sky: from enormous systems like Complex C, subtending
an angular area of roughly $90^{\circ} \times 20^{\circ}$ \citep{wakker91}, to
subdegree-scale compact high-velocity clouds, some of which are
unresolved even with large single-dish radio telescopes
\citep{braun99,putman02}.  As their name suggests, the radial velocities of
HVCs relative to the local standard of rest ($\left| v_{\rm LSR}\right| >
100$\,km\,s$^{-1}$) are inconsistent with Galactic rotation, implying they
reside either in the extended Galactic halo or at extragalactic (Local Group)
distances.  HVCs appear to be a common feature around spiral galaxies, with M31
\citep{thilker04}  and M83 \citep{miller09} hosting what may be analogs to the
largest HVCs seen near the Galaxy.

Numerous physical scenarios for the formation of HVCs have been proposed,
including the cooling component of a supernova-driven ``Galactic fountain,''
\citep{bregman80,booth07,kwak09}, inflows of neutral gas condensing from the
local intergalactic medium \citep{putman06,sommer06,keres09}, or the gaseous
signatures of the ``missing'' dark matter subhalos around the Galaxy
\citep{moore99,giovanelli10}.  Oddly, unlike recently-discovered, very low-mass
satellite galaxies which appear to harbor both stars and dark matter
\citep[e.g.,][]{strigari08} but little or no gas \citep{grcevich09}, searches
for stellar populations associated with HVCs have only succeeded in placing
upper limits \citep{willman02,siegel05,simon06,hopp07}. Similarly, aside from
one case \citep{richter99}, HVCs appear to contain little or no molecular gas
\citep{wakker97,akeson99,combes00}.  The detection of associated ionized gas
\citep[up to O\,{\sc vi};][]{sembach03,shull09}, but not the very high ions
tracing $10^6$K gas \citep{williams06}, supports the notion that HVCs may be
related to nonequilibrium cooling processes, either from the IGM or a Galactic
fountain.  

Thus, although the most extended HVCs appear to be located relatively near the
Galaxy \citep{wakker08,lockman08,smoker11}, they may nonetheless be connected
to extragalactic phenomena. Complex C, for example, lies at a distance of $\sim
10$\,kpc \citep{wakker07,thom08} and exhibits a low but inhomogeneous
metallicity \citep{gibson01,tripp03,collins07} that may be indicative of a
combination of infalling primordial gas and outflowing enriched material.
While \emph{compact} HVCs (CHVCs) were initially hypothesized to be faint Local
Group denizens \citep{braun99}, evidence is mounting that most of them simply
represent the smallest fragments of the ``normal'' HVC population. For
instance, \citet{maloney03} concluded that, due to their large inferred masses and
inconsistent velocity widths, CHVCs as a population are more likely to be
associated with the Galactic halo than the Local Group; additionally, searches
for similar clouds in nearby Local Group analogs indicate that CHVCs are probably
clustered within 90\,kpc of the Galaxy, and massive intragroup \hi\
clouds are relatively rare \citep{pisano04,pisano07,chynoweth09}. The lack
of large \hi\ clouds beyond 50 projected kpc from M31 \citep{westmeier08} further suggests
that most of the HVCs around our analogous spiral Galaxy do not lie far beyond the halo.
\citet{putman11} recently found that at least 35\% of CHVCs exhibit head-tail
morphologies \citep[suggesting interaction with a diffuse warm-hot Galactic
halo medium;][]{williams06,williams07,bregman09} and most of these lie near
larger HVC complexes.  In certain rare cases, though, CHVCs which are \emph{not} 
associated with larger-scale H\,{\sc i} structure may be candidates for low-luminosity 
dwarf galaxies.  

The reported detection of dust in Complex C by \citet[][hereafter 
M05]{miville05} provides a
novel way to test the connection between HVCs and CHVCs.  If CHVCs are simply
scaled-down versions of Complex C at similar distances, their dust emission
properties should be similar.  On the other hand, if certain CHVCs differ
fundamentally from the non-compact HVCs (e.g., in their metallicities or
Galactocentric distances), this would manifest itself through correspondingly
different dust emission properties.  M05 inferred the dust content of Complex C
by cross-correlating the far-IR surface brightness in
the \emph{Spitzer Extragalactic First-Look Survey} with \hi\ observations in the
same field \citep{lockman05}; however, a subsequent analysis by \citet{peek09}
using more sensitive Arecibo \hi\ measurements and IRAS fluxes
did not confirm the M05 detection in Complex C (but, interestingly, detected 
dust in HVC Complex M). New longer-wavelength data from the \planck\ satellite
\citep{planck11a} may have finally found dust in Complex C, albeit at the 
$\sim 3\sigma$ level \citep[][hereafter Planck11]{planck11b}. 
In short, the prevalence of dust in HVCs is still essentially unknown, yet
this information is crucial to our understanding of both HVCs themselves
and their significance in the Milky Way and Local Group.

With much smaller sizes and often high peak \hi\ column densities, CHVCs
may allow the \emph{direct} detection of dust emission; such detections
(or strong upper limits thereupon)
can therefore provide insight to the nature of these mysterious objects.  
Here we present a targeted \emph{Spitzer-MIPS} search for dust emission
in three CHVCs: one with high-resolution radio data revealing several
dense arcminute-scale knots, and two other high-column density CHVCs
from the \hi\ Parkes All-Sky Survey (HIPASS) catalog \citep{putman02}.  This paper is organized as follows: in
Section~\ref{sec_data} we discuss the CHVC sample and the \emph{Spitzer}
observations; Section~\ref{sec_measure} describes the flux measurements from
the \emph{Spitzer} images, in Section~\ref{sec_discussion} we discuss possible
interpretations of the \emph{Spitzer} observations (and caveats thereto),
with the overall results briefly summarized in Section~\ref{sec_summary}.

\section{Sample and Observations} \label{sec_data}
Due to its extremely compact, ``clumpy'' morphology as revealed by
the high-resolution radio maps of BW04, and their conclusion that
this is potentially a good candidate for a ``missing'' dark matter
minihalo, we selected CHVC289 as our primary
follow-up target.  This object exhibits high peak \hi\ column 
densities in all its five knots ($N_H>10^{20}$\,cm$^{-2}$) in the 
high-resolution maps, even though its properties appear unremarkable in
single-dish observations (peak $N_H\sim 2.9\times 10^{19}$\,cm$^{-2}$
as observed with Effelsberg), having a relatively small total flux compared 
to other CHVCs due to its small size.  

To complement this observation, we selected two other CHVCs from the HIPASS
catalog of \citet{putman02}: CHVC $319.1-78.8+215$  and CHVC $147.8-82.4-268$
(hereafter referred to as CHVC319 and CHVC148 respectively).  These were
specifically chosen because of their very high peak column densities averaged
over the Parkes beam ($N_H>5\times 10^{19}$\,cm$^2$), and compact sizes
(semimajor axes less than 0.2 degrees).  In addition to the HIPASS column
densities and sizes, higher-resolution observations of CHVC148 were obtained by
\citet{westmeier05} on the 100\,m Effelsberg single-dish telescope and by
\citet{deheij02} on the Westerbork interferometer.  Although the detailed
(arcminute-scale) structures of these CHVCs are unknown due to the lack of
high-resolution imaging comparable to that obtained by BW04 on CHVC289, their
HIPASS column densities are over four times higher than that of CHVC289, making
them good candidates for objects with high-column density \hi\ knots that may
harbor dust.

All three of these CHVCs were observed in Cycle 3 with the 
\emph{Spitzer-MIPS} $24\mu$m, $70\mu$m, and $160\mu$m channels
(MIPS1, MIPS2, MIPS3 respectively; see Table~\ref{tab_log} for
a summary of the observations), and the data were processed using
standard techniques given in the \emph{Spitzer} Data Analysis
Cookbook\footnote{\url{http://ssc.spitzer.caltech.edu/dataanalysistools/cookbook/}}.  
Because of its extreme compactness and precisely-constrained \hi\ knot
positions, CHVC289 was observed with a single MIPS large-field AOT in MIPS1 and
MIPS3.  Since half of the MIPS2 array is unusable, in this band we used two
separate observations offset by 160\arcsec\ to ensure complete coverage of the
CHVC.  These observations were designed to provide a full 5\arcmin\ by
5\arcmin\ image of the central portion of CHVC289 in all three bands, though
due to an inoperative MIPS3 readout there are two narrow strips of missing data
in the $160\mu$m image.  The nominal integration time per pixel in channels 1,
2, and 3 are 648, 629, and 52 seconds respectively.

The other two CHVC positions are only known to within a few arcminutes and
their angular sizes are larger (albeit with significantly higher column
densities than CHVC289); we thus instead obtained MIPS scan maps of the regions
surrounding these two CHVCs, but to lower sensitivity since their high column
densities should correspond to stronger dust emission.  The scan maps were
designed to cover $24\arcmin\times 30\arcmin$ regions with $1/4$ array
(80\arcsec) offsets to fill in the gaps caused by the bad MIPS3 detector
pixels; this area is sufficient to fully cover CHVC148 and CHVC319 with at
least 2 overscans in each band.  Basic characteristics of these observations
are also listed in Table~\ref{tab_log}.

\section{Measurements} \label{sec_measure}
\subsection{Mid-IR fluxes and upper limits}
\begin{figure}
\plotone{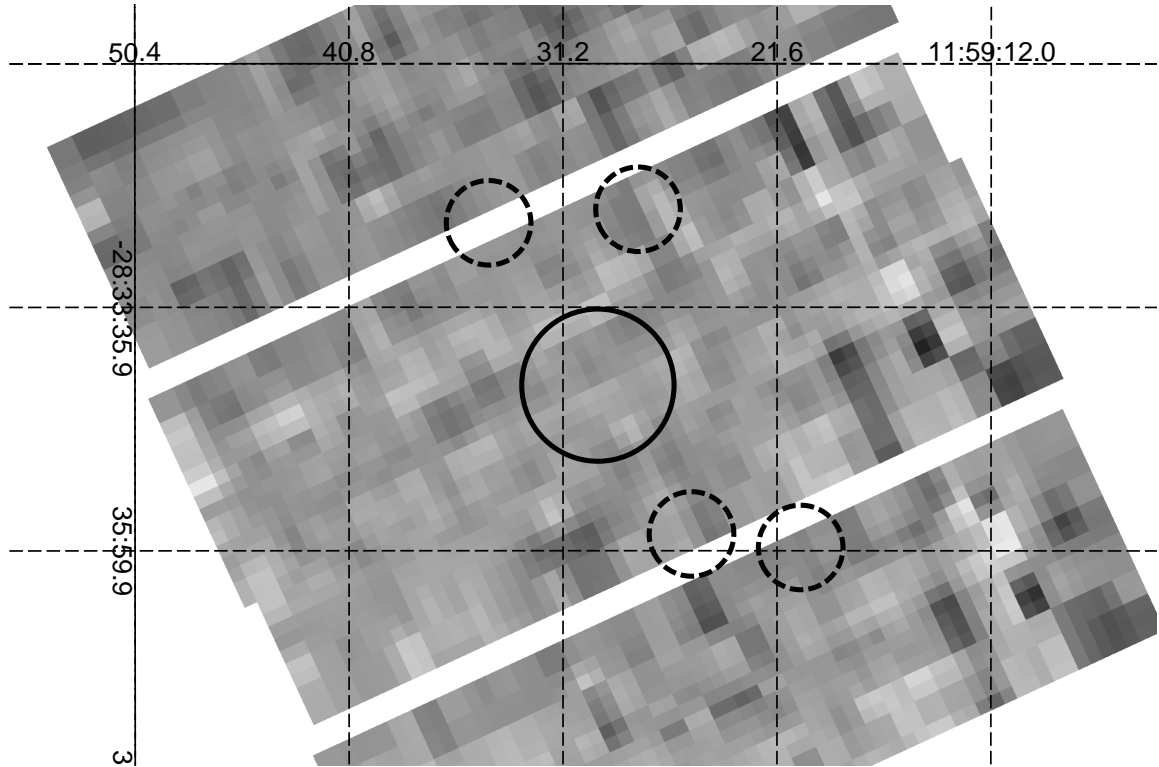}
\caption{Spitzer-MIPS $160\mu$m image of CHVC289 with equatorial (J2000) 
coordinates overlaid.  Circles indicate the positions
and approximate sizes of the five cores reported by BW04 (with dashed
circles denoting unresolved \hi\ emission); the gaps in coverage are a
result of bad detector pixels in this band.  Although any dust emission
from this CHVC is expected to be brightest at $160\mu$m, no significant 
emission is seen at any of the clump positions (though three of the clumps fall 
on or near image gaps).
\label{fig_chvc289}}
\end{figure}

\begin{figure*}
\plottwo{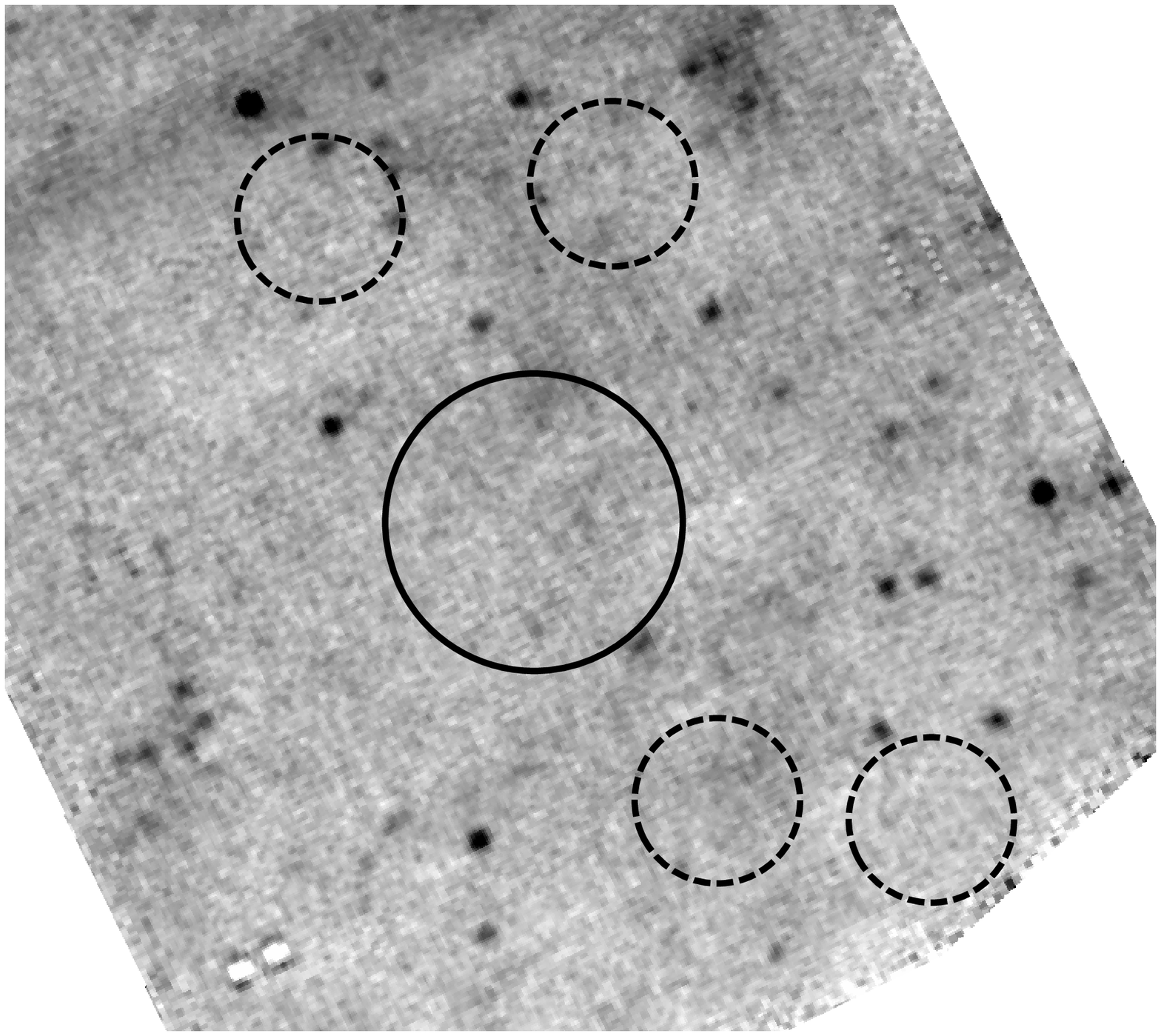}{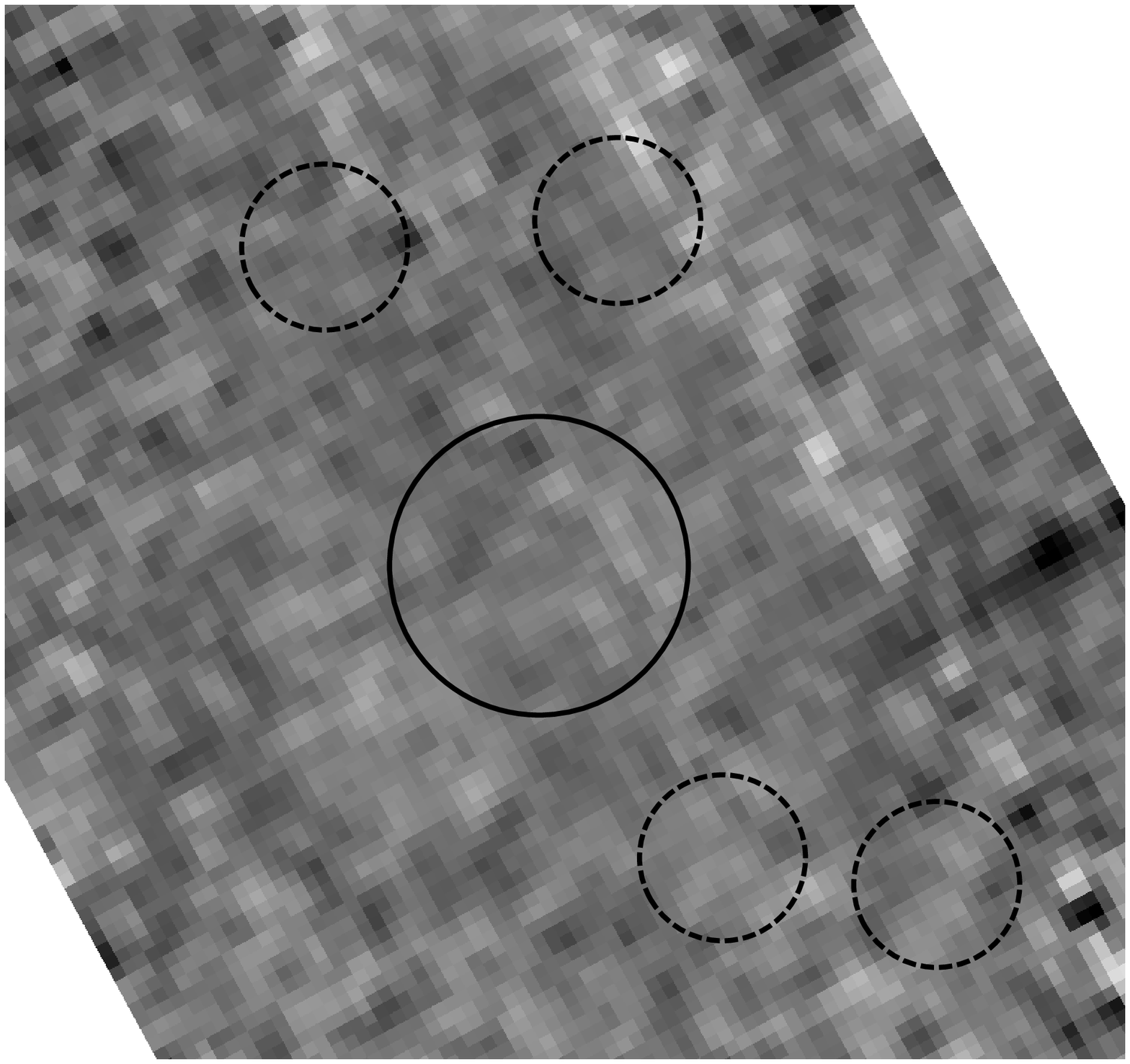}
\caption{MIPS $24\mu$m (left panel) and $70\mu$m (right panel) images
of CHVC289.  Clump positions and sizes are shown with circles as in 
Figure~\ref{fig_chvc289}.  Dust emission is expected to be weaker
in these bands than in $160\mu$m.  Even though the $24\mu$m image shows 
a number of unresolved point sources, these are not correlated with the 
clump positions and are most likely background galaxies.
\label{fig_chvc289_ch12}}
\end{figure*}

For CHVC289, we employ the known positions and sizes of the \hi\ 
clumps measured by BW04 with the \emph{Australia Telescope Compact
Array} (ATCA), shown overplotted on the MIPS images in Figures
\ref{fig_chvc289} and \ref{fig_chvc289_ch12}.  The central, highest-column 
density clump was resolved by 
ATCA with a diameter of 90\arcsec.  This is roughly $1.5-2$ times larger
than the MIPS $160\mu$m point response function and therefore
(provided the dust and \hi\ densities are correlated) the mid-IR dust 
emission from this clump should similarly be resolved by all MIPS 
channels.  The four less-dense clumps are
unresolved by ATCA (with sizes less than 50\arcsec); unfortunately, these
fall either partly or entirely on the bad-pixel gaps in the 
MIPS3 image, so we only consider the dense central clump here.

Since the emission is nominally extended rather than a point-source,
we measure the flux of the central CHVC289 clump in each of the MIPS 
images with a 90\arcsec\ diameter circular aperture (i.e., approximately equal
to the \hi\ size measured with ATCA).  The
uncertainty on the flux is determined with the ``empty aperture method,'' 
where the same flux measurement is repeated for many randomly placed 
apertures on source-free regions of 
each image; the resulting dispersion in the flux measurements represents
the uncertainty in the background levels, while the median provides an
estimate of the background levels in the image.  In none of the three MIPS 
bands is the flux in the CHVC289 cores significant; we therefore compute
$2\sigma$ upper limits in all bands assuming the measured uncertainties.

\begin{figure*}
\includegraphics[width=9.0cm]{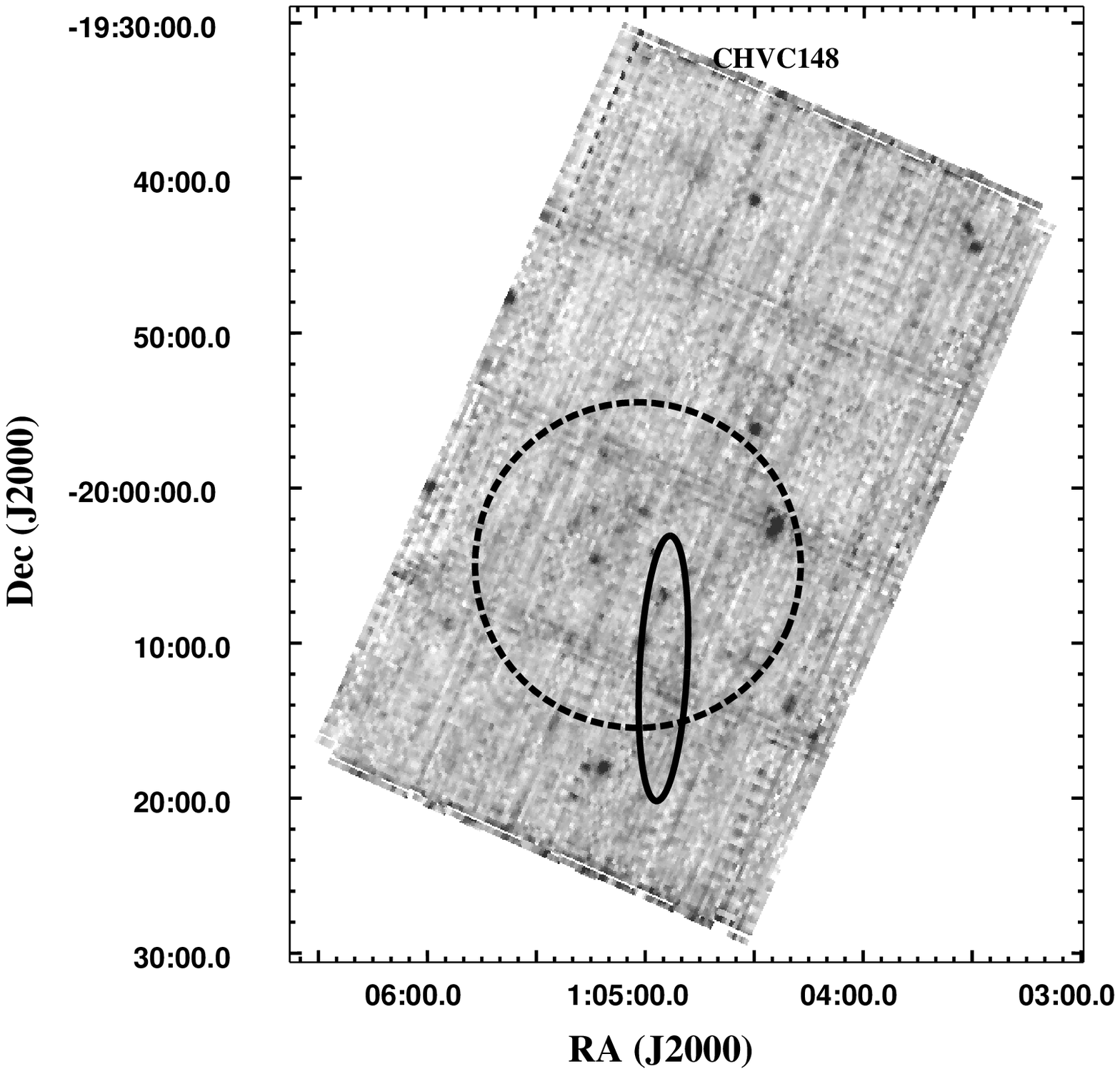}
\hspace{-1.0cm}
\includegraphics[width=10.4cm]{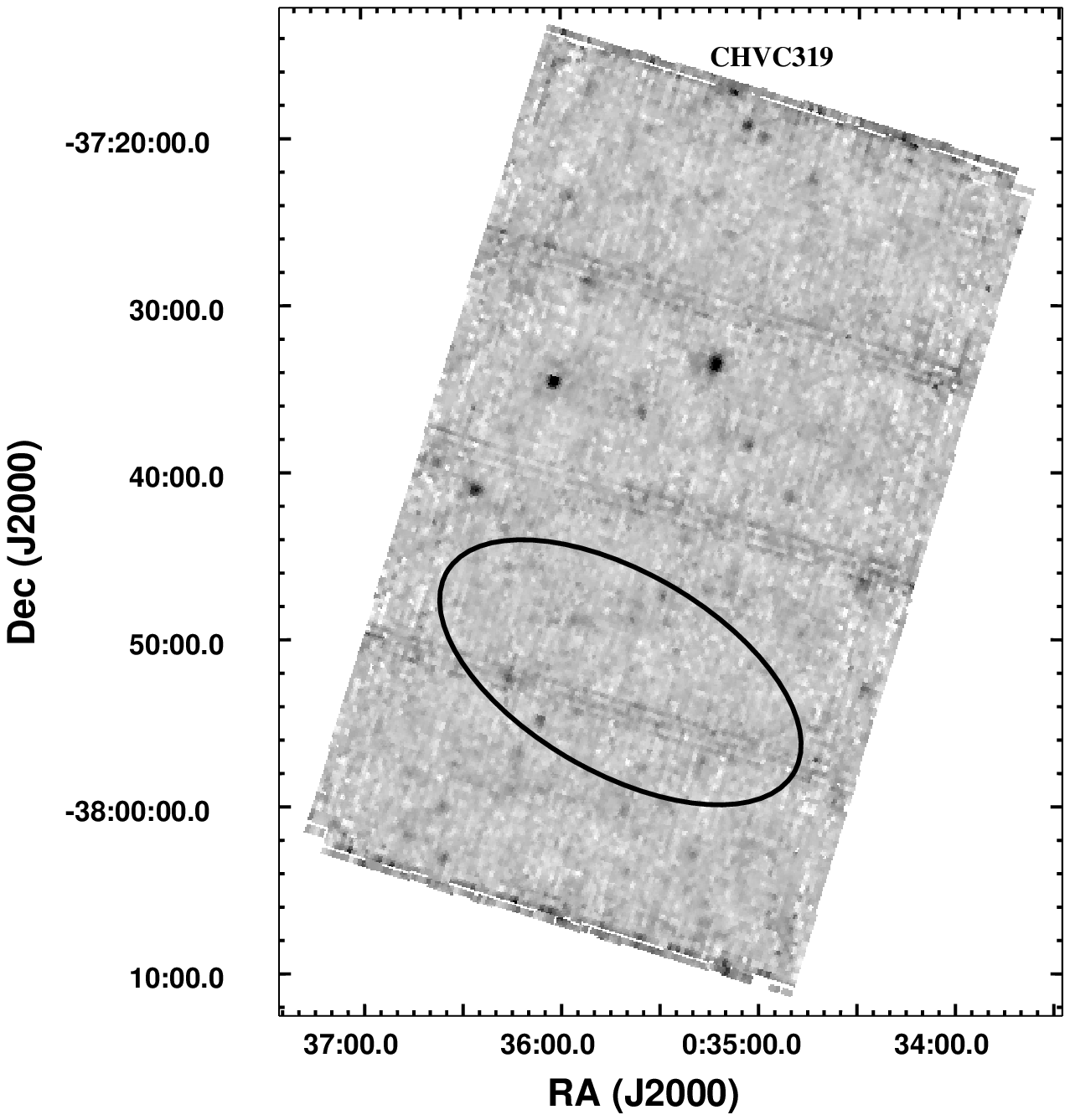}
\caption{\emph{Left panel:} MIPS $160\mu$m scan maps of the region around 
CHVC148.  The dashed circle shows the approximate position and beam
size of the \citet{westmeier05} Effelsberg observation, while the
narrow solid ellipse denotes the compact core reported by \citet{deheij02}
containing 4\% of the CHVC's flux (note that this represents their 
synthesized Westerbork beam, not the actual size and shape of the core).
While many compact sources fall within the CHVC region, all correspond
to background galaxies.
\emph{Right panel:} Same, but for CHVC319; here, the ellipse denotes
the CHVC size and shape reported in the HIPASS catalog \citep{putman02}.  
No significant extended emission is apparent in either of these CHVCs.
\label{fig_otherchvc}}
\end{figure*}

The small-scale \hi\ structures and positions of CHVC148 and CHVC319 are 
less well-constrained: the estimated size of CHVC319 in the HIPASS
catalog is given as $0.2^{\circ}\pm 0.1^{\circ}$ \citep{putman02}, while 
CHVC148 has a FWHM of $\sim 20\arcmin$ \citep{westmeier05}.
The uncertainty ellipses of these clouds take up a substantial
fraction of the MIPS field; thus, the empty aperture method cannot be
used to estimate flux uncertainties in these cases.  Note that although
\citet{deheij02} report a compact \hi\ core in CHVC148, this comprises only
4\% of the total \hi\ flux from this CHVC with a peak column density
of $\sim 10^{19}$cm$^2$, or roughly an order of magnitude less than
the central concentration in CHVC289; any dust emission from this
core is therefore likely to be negligible.  The MIPS $160\mu$m images 
of both of these CHVC regions are shown in Figure~\ref{fig_otherchvc},
with the positions and beam sizes overplotted.

No extended $160\mu$m
emission is evident within the CHVC areas, but a number of point sources 
are visible, particularly in the CHVC148 region.  We ran
\emph{Source Extractor} \citep{bertin96} on the $160\mu$m image to detect
point sources and accurately measure their positions,
and checked the other two MIPS images for counterparts. $24\mu$m sources
are clearly visible at the position of all $160\mu$m objects
in CHVC148, and a search of the 
\emph{NASA-IPAC Extragalactic Database}\footnote{The NASA/IPAC Extragalactic 
Database (NED) is operated by the Jet Propulsion Laboratory, California 
Institute of Technology, under contract with the National Aeronautics and 
Space Administration.} 
reveals that each of these sources lies within $30\arcsec$ (i.e., roughly 
within the $160\mu$m point-response function) of a relatively 
bright galaxy.  We therefore conclude that these compact mid-IR sources 
are background galaxies rather than dust clouds intrinsic to the CHVCs,
and both of these CHVCs therefore lack significant extended or compact
dust emission.

% 160um: 0 +- 21 (90"x1.5 aper)
%	 0 +- 13 (90" aper)

\subsection{The CHVC289 mid-IR spectral energy distribution}
As noted in the previous section, we were only able to derive upper
limits on the dust emission in CHVC289 in all three \spitzer\ bands.
With previously-measured correlation coefficients between \hi\ 
column density and dust emission, we can 
predict the mid-IR surface brightnesses that the densest core of CHVC289
\emph{should} exhibit if its dust properties are similar
to HVC Complex C or intermediate-velocity clouds (IVC).
Here we employ measurements made by two independent groups: M05 in the \emph{Spitzer
Extragalactic First Look Survey}, and the ``Draco'' field analyzed by Planck11, both of
which cover small parts of Complex C.
These analyses claimed a significant correlation of mid-IR emission with \hi\
in Complex C, assuming a simple proportionality 
($I_\nu\sim \alpha_\lambda N_H$ in M05, $I_\nu\sim \epsilon_\nu N_H$ in Planck11; 
both define the emissivities $\alpha_\lambda$ and $\epsilon_\nu$ in units of 
MJy~sr$^{-1}$~($10^{20}$~cm$^{-2}$)$^{-1}$). Furthermore, in their fitting,
both allowed for independent mid-IR emission from known local, IVC, and HVC \hi\ 
velocity components. Planck11 only include one IVC component in their fits while
M05 consider two; for clarity, in our comparison to M05, we only consider their
``IVC1'' and ``HVC'' components.

\begin{figure}
\plotone{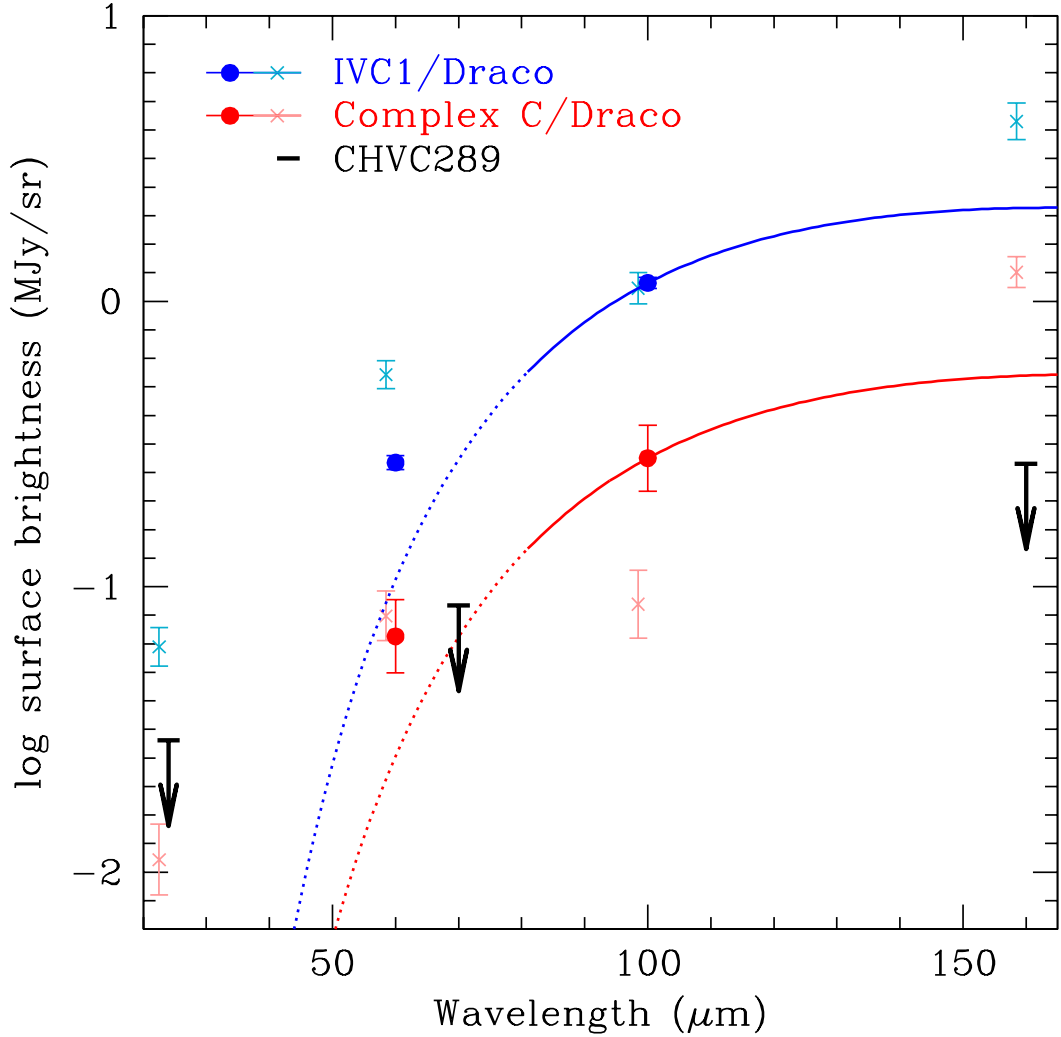}
\caption{Surface brightness as a function of wavelength for CHVC289
($2\sigma$ upper limits; black arrows), compared to the values expected if
CHVC289's dust emission properties are similar to the HVC (red points and
lines) and IVC (blue points and lines) components in the HVC Complex C/\planck\
Draco field. This comparison includes IR-\hi\ correlation coefficients reported
by M05 (light-colored crosses), and \emph{IRAS/Planck} (dark circles;
Planck11).  As noted in Section~\ref{sec_consistency}, the M05 uncertainties
are likely underestimated.  Lines denote $100\mu$m-normalized, modified black
body fits to the $100-850\mu$m \emph{IRAS} and \planck\ data; note that dust
emission at shorter wavelengths is dominated by non-equilibrium processes and
therefore not expected to fit the black body curve at $60\mu$m and $24\mu$m.
The $160\mu$m surface brightness upper limit is a factor of $\sim 2$ below the
predicted emission from \planck, and IVC-like dust emission is strongly ruled
out in all three bands. This discrepancy could be due to lower dust content in
CHVC289 than Complex C, lower temperature, or a combination of both. 
\label{fig_flux}}
\end{figure}

In figure~\ref{fig_flux} we show these expected surface brightnesses for the
central \hi\ core of CHVC289, assuming the \citet{bruns04} column
density of $1.58\times 10^{20}$\,cm$^{-2}$.  Estimates for the measured 
Planck11 \emph{IRAS} and M05 \emph{IRAS}/\emph{MIPS} IR-\hi\ correlations
are shown as points. Additionally, we estimate modified blackbody 
spectra\footnote{The modified blackbody function employed by Planck11 scales
as $\epsilon_\nu \sim \nu^\beta B_\nu(T)$, where
$\beta$ and $T$ are the free parameters and $B_\nu(T)$ is the standard Planck
function.}
based on the 100, 350, 550, and $850\mu$m measurements in the 
Planck11 Draco field (Table 2). These curves are normalized to the IRAS $100\mu$m
flux, and the IRAS $60\mu$m point is not included in the fit since dust
emission at high frequencies is dominated by nonequilibrium processes.

We convert our mid-IR flux limits for the central CHVC289 core to surface
brightness upper limits by assuming that the flux is distributed evenly over
the 90\arcsec\ measurement aperture. Since this is comparable to the size
(FWHM) of the \hi\ emission itself, in essence we are comparing the predicted
and measured \emph{average} surface brightnesses in the central CHVC289 core;
due to the effects of beam smearing and the lack of a $160\mu$m detection, this
is a reasonable approximation.  These derived upper limits for all three bands
are overplotted in \ref{fig_flux}.  Although the two bluer MIPS upper limits
($24\mu$m and $70\mu$m) are more or less consistent with the predicted emission
for Complex C-like dust, the $160\mu$m upper limit is a factor of $\sim 4$
below the prediction from M05, and a factor $2$ below the Planck11 blackbody
curve; the two redder MIPS bands strongly rule out IVC-like emission (and our
$24\mu$m measurement appears inconsistent with the expected flux from M05).
Unfortunately, without a detection in any of the bands, we are unable to place
limits on the dust temperature in CHVC289.

\section{Discussion} \label{sec_discussion}
\subsection{Consistency between M05 and Planck11} \label{sec_consistency}
As alluded to in Section 1, the presence of dust in HVC Complex C has been a
somewhat contentious topic. In large part this stems from the presence of
multiple Galactic components along the line of sight (including local ISM, one
or more intermediate-velocity components, and Complex C itself) which vary in
column density, velocity, and even composition across the enormous area spanned
by Complex C. Mid-IR emission, by contrast, lacks the detailed velocity
information of \hi, so 2D spatial correlations must take into account
contributions from each distinct component, with each potentially exhibiting a
different dust fraction and temperature. As a result, there can be substantial
degeneracies between the various measured parameters, and properly accounting
for uncertainties is paramount.

M05 first reported a significant detection of dust (with $\epsilon_{160}=0.8\pm 0.1$,
nominally an $8\sigma$ result in this band) over a $\sim 5$\,deg$^2$ field, using
a $\chi^2$ minimization technique; both calibration and statistical uncertainties
were included. However, as pointed out by \citet{peek09}, spatial variations
in the dust-to-gas ratio of the ``local'' component (which by far dominates
the IR emission) can introduce spurious signals for weaker components like
the HVC. Since this possibility was not taken into account by M05, and since
their analysis does not account for position-to-position statistical
variations (like the empty aperture method we employ, or the ``displacement
map'' technique described by \citealt{peek09}), it appears likely that M05's
uncertainties are underestimated (especially for the HVC component).

Given this, plus the fact that M05 and Planck11 analyzed different parts
of Complex C, the discrepancies between some of the M05 and Planck11 data 
points in Figure~\ref{fig_flux} are not surprising.  In particular it is
notable that the quoted errors on the $60\mu$m and $100\mu$m fluxes are comparable
between Planck11 and M05, despite the factor $\sim 5$ larger angular area
analyzed by Planck11. On the other hand, Planck11 do not include data at $24\mu$m,
and at $160\mu$m we can only interpolate their modified blackbody
fits. Due to their far wider spatial and spectral coverage, we take the Planck11
as the most reliable comparison to our data; however, for completeness we
include comparisons to the M05 points as their \spitzer\ measurements are
more analogous to ours. At worst, if the M05 $160\mu$m point is taken as
an upper limit, our data provide a nearly order-of-magnitude more stringent limit, 
and also lie a factor $\sim 2$ below the expectation from Planck11.

\subsection{Are CHVCs a distinct, dust-poor population?}
None of the CHVCs studied here show evidence 
for the presence of dust in our deep \emph{Spitzer-MIPS} imaging, despite 
their relatively high column densities.  CHVC289 in particular, with its 
strong central core resolved with ATCA, should have
been confidently detected at the $4-8\sigma$ level in the $160\mu$m band
(and perhaps marginally at $70\mu$m) 
if its dust temperature and dust-to-gas ratio are similar to HVC 
Complex C.  As shown in Figure~\ref{fig_flux}, the $160\mu$m 
upper limit is incompatible with a similar origin, lying a factor of
$\sim 2$ below the flux expected from the Planck11 measurements.  Even more striking
is the discrepancy in the two redder bands between CHVC289 and IVC1.  
Intermediate-velocity
clouds are thought to reside closer to the Galactic disk, perhaps
resulting from outflowing material due to star formation or other
processes;  however, the limits on CHVC289 are a factor of $2-10$ below
the IVC emission level predicted by either M05 or Planck11.  
The dust-gas correlation for IVC2 as
reported by M05 is not shown in Figure~\ref{fig_flux}, but lies in 
between Complex C and IVC1. The extreme compactness of CHVC289, along
with its lack of detected IR emission, suggest that it
represents a different type of system than either Complex C or the IVCs.

However, with the current data we cannot ascertain whether this difference
is intrinsic to CHVC289, or simply due to a greater distance from the
Galactic plane (hence weaker ambient radiation field and lower dust
temperature/emissivity).   The upper limit to CHVC289's $160\mu$m 
surface brightness is a factor 4 below the expectations from M05, who 
estimate a 
dust temperature in Complex C of $10.7$\,K (though the uncertainties on this
quantity are likely large, as well).  Assuming black body emission,
this discrepancy in flux is consistent with CHVC289 having a 
lower temperature, $<9.2$\,K, which in turn means (if the temperature difference
depends entirely on Galactocentric distance and the dust heating scales
as $\sim 1/R_G^2$) that CHVC289 could in principle be only 
marginally ($\gtrsim 10$\%) farther away than Complex C's average
distance of 10\,kpc \citep{wakker07,thom08}. An even smaller difference in
distance could explain the discrepancy between CHVC289 and the better-constrained
Planck11 measurements.  Of course, this rough estimate assumes 
that these two clouds have identical dust properties aside from their temperatures;
in reality other variables are likely to be in play (e.g., dust distribution
and dust-to-gas ratio).  A robust determination of the physical
conditions in CHVC289 thus evades us, since the current data cannot
directly constrain the dust temperature of this cloud.

Similarly, no mid-IR emission is seen in either of the other two systems 
presented here, CHVC148 and CHVC319. However, this result is weaker since
the small-scale structures of these CHVCs are essentially unknown.
These objects were initially chosen for \emph{MIPS} follow-up due
to their compact (single-dish) sizes and far higher peak column densities
than CHVC289; given that ultra-compact, high-density cores in CHVC289 were
found under interferometric scrutiny by BW04, correspondingly higher-density
(or more numerous) cores may also exist in CHVC148 and CHVC389.  
Nonetheless, neither obvious clumps nor diffuse haze in the mid-IR
is visible within either of these systems.  \citet{deheij02} did in fact
find a ``core'' (albeit with much lower resolution than the BW04
observations) in CHVC148, but this core comprises only 4\% of the
total HVC flux.  Higher-resolution \hi\ observations of these
systems would allow a more detailed comparison with the MIPS data;
however, since no sources are seen in the mid-IR, these comparisons
will likely produce only weak upper limits (as with CHVC289).

\subsection{Caveats}
This analysis partly hinges on a comparison of the mid-IR emission
in the CHVCs to estimates of the dust content of Complex C, which
is still uncertain even with the Planck11 data; the primary reason
for this is the weakness of the Complex C dust emission and dominant
foreground \hi/dust components.
Moreover, as M05 point out, despite the evidence that Complex C contains 
dust, no molecular gas  has been detected thus far; either through 
H$_2$ absorption \citep{murphy00,richter01} or $^{12}$CO emission 
\citep{dessauges07}.  Since dust and molecular gas elsewhere in
the Galaxy are strongly correlated, this appears somewhat incongruous.
M05 suggest that the dust in Complex C may instead be confined
to cold, compact cores with the surrounding warmer gas comprising
most of the \hi; however, our observations strongly rule out such
emission in CHVC289's compact cores.  On the other hand, this may 
further highlight that CHVC289 and Complex C are different
in some elementary way besides their angular sizes.

Although the central clump of CHVC289 is resolved with ATCA, it is
only marginally so: the inferred clump size is 90\arcsec, compared
to the $112\arcsec\times 36\arcsec$ beam reported by \citet{bruns04}.
The morphology of the clump could actually be more complex, consisting
of several denser clumps spread over the 90\arcsec area.  Since we assumed
constant \hi\ column density and dust emission over this area, the
presence of such clumps could ``hide'' emission in higher (but
still undetected) surface brightness regions, implying that our upper
limits could in principle be too low.  However, such effects are
unlikely to account for the large (factor of $\sim 2$ at $160\mu$m)
discrepancy between the CHVC289 and Complex C emission properties,
especially given the large \hi\ column density measured over the
same 90\arcsec\ area.

Since only upper limits in all three bands can be calculated for CHVC289, 
it is not possible to compare the shape of the putative dust SED
with the other systems (Complex C and the IVCs).  If CHVC289 has
a substantially flatter dust SED (e.g., due to a different grain size
distribution or higher temperature), this may also account for its
lower $160\mu$m flux.  On the other hand, such scenarios would be
necessarily contrived (e.g., CHVC289 would need to be both hotter
and very dust poor), and the overall conclusion would remain: 
it is a fundamentally different system than the ones studied
by M05 and Planck11.

\section{Summary and Outlook} \label{sec_summary}
Using deep \emph{Spitzer-MIPS} observations of three CHVCs, we have placed
limits on the dust emission from these enigmatic objects, ruling out similar
physical conditions in HVC Complex C and the most compact object in our sample
(CHVC289). Notably, the $160\mu$m surface brightness limit falls a factor of 2
below that inferred via \emph{Planck} measurements for dust in HVC Complex C,
and are even more discrepant with IVC measurements, indicating that CHVC289 is
either farther away or exhibits a lower dust-to-gas ratio than these other two
object categories.  CHVC289 (and by extension other CHVCs) may therefore reside
farther from the Galactic disk or exhibit more ``primordial'' compositions.
Alternatively, if all (C)HVCs exhibit similar physical conditions and the
M05/Planck11 dust detections in Complex C are spurious, our CHVC289 measurement
places the most stringent upper limit on HVC dust yet. Unfortunately, since all
we have are upper limits, the current data do not allow us to distinguish
between these possibilities.

No extended dust emission is seen toward the other two CHVCs (CHVC148 and
CHVC389) observed with \emph{MIPS}; however, their small-scale \hi\
distributions are poorly constrained.  Since their total column densities are
much higher than CHVC289 with comparable radii measured from single-dish
observations, high-resolution \hi\ maps of these two clouds might allow
stronger constraints to be placed on their dust emission with the existing
\emph{MIPS} data.  A robust study of the dust content over the full CHVC
population (and comparison to large-scale HVCs) will require of a much larger
sample than the three presented here. In particular, the \planck\ mission
\citep{planck11a}, with its full-sky coverage, sensitivity to cold dust at
submillimeter wavelengths, and angular resolution well-matched to CHVC sizes,
will prove integral to this aim. 

% 2-sig upper limits on CHVC289 emission (90" diam aperture)
% 160um = 26.70, 70um = 34.1, 24um = 465
% converted to MJy/sr: 160=0.27, 70=0.086, 24=0.029 

\acknowledgments
We are grateful to the anonymous referee for comments which substantially improved
the manuscript,
and to the \emph{Spitzer} team for their efforts with this remarkable
instrument.  Financial support for this work was provided by NASA through 
award number 1289595 issued by JPL/Caltech.
R.J.W. acknowledges additional support from the 
Netherlands Organization for Scientific
Research (NWO), the Leids Kerkhoven-Bosscha Fonds, and NSF grant
AST-0707417. This research has made use of the NASA/IPAC Extragalactic 
Database (NED) which is operated by the Jet Propulsion Laboratory, 
California Institute of Technology, under contract with the National 
Aeronautics and Space Administration.

\clearpage

\begin{deluxetable*}{lccccccccc}
\tablecolumns{10}
\tablewidth{500pt}
\tablecaption{\emph{Spitzer} observation log \label{tab_log}}
\tablehead{
\colhead{Target} &
\colhead{RA} &
\colhead{Dec} &
\colhead{$t_{\rm exp,24}\tablenotemark{1}$} &
\colhead{$t_{\rm exp,70}$\tablenotemark{1}} &
\colhead{$t_{\rm exp,160}$\tablenotemark{1}} &
\colhead{$\sigma_{24}$\tablenotemark{2}} &
\colhead{$\sigma_{70}$\tablenotemark{2}} &
\colhead{$\sigma_{160}$\tablenotemark{2}} &
\colhead{Notes}\\
\colhead{} &
\colhead{(J2000)} &
\colhead{(J2000)} &
\colhead{(s)} &
\colhead{(s)} &
\colhead{(s)} &
\colhead{(MJy/sr)} &
\colhead{(MJy/sr)} &
\colhead{(MJy/sr)} &
\colhead{} 
}

\startdata
CHVC289 &$11:59:30$ &$-28:34:32$ &648 &629 &52 &0.073 &0.45 &1.1 &single pointing\\
CHVC148 &$01:05:06$ &$-20:09:00$ &168 &84 &17 &0.18 &0.39 &0.54 &MIPS scan map\\
CHVC319 &$00:35:42$ &$-37:52:00$ &168 &84 &17 &0.068 &0.35 &0.38 &MIPS scan map
\enddata
\tablenotetext{1}{ Exposure times are seconds per pixel as estimated from the Astronomical Observation Requests (AORs) }
\tablenotetext{2}{Median pixel-to-pixel RMS values in the regions covering  
each CHVC}
\end{deluxetable*} 

\end{document}